\documentclass[lettersize,journal]{IEEEtran}
\usepackage{cite,color}
\usepackage{amsmath,amssymb,amsfonts}
\usepackage{algorithmic}
\usepackage{algorithm}
\usepackage{array}
\usepackage{url}
\usepackage[normalem]{ulem}
\useunder{\uline}{\ul}{}
\usepackage{float}
\usepackage{subfigure,multirow,bm,stfloats}
\usepackage{textcomp}
\usepackage{amsthm}

\usepackage{graphicx}
\usepackage[table]{xcolor}

\begin{document}

\title{RME-GAN: A Learning Framework for Radio Map Estimation based on Conditional Generative Adversarial Network}

\author{Songyang~Zhang, Achintha~Wijesinghe,
	and~Zhi~Ding,~\IEEEmembership{Fellow,~IEEE}
        % <-this % stops a space
\thanks{* S. Zhang, and A. Wijesinghe have the same contribution to this work.}% <-this % stops a space
\thanks{This work was supported in part by the National Science Foundation under Grant 2029848 and Grant 2029027.}
\thanks{S. Zhang, A. Wijesinghe and Z. Ding are with Department of Electrical and Computer Engineering, University of California, Davis, CA, 95616. (E-mail: sydzhang@ucdavis.edu, achwijesinghe@ucdavis.edu, and zding@ucdavis.edu).}}

% The paper headers
\markboth{Journal of \LaTeX\ Class Files,~Vol.~14, No.~8, August~2021}%
{Shell \MakeLowercase{\textit{et al.}}: A Sample Article Using IEEEtran.cls for IEEE Journals}

%\IEEEpubid{0000--0000/00\$00.00~\copyright~2021 IEEE}
% Remember, if you use this you must call \IEEEpubidadjcol in the second
% column for its text to clear the IEEEpubid mark.

\maketitle

\begin{abstract}
Outdoor radio map estimation is an important {tool} for network planning and resource management in modern Internet of Things (IoT) and cellular systems. Radio map describes spatial signal strength distribution and provides  
network coverage information. A practical goal is to estimate
fine-resolution radio maps from sparse radio {strength} measurements. 
However, non-uniformly positioned measurements and 
access obstacles can make it difficult for accurate radio map estimation (RME)
and spectrum planning in many outdoor environments.
In this work, we develop a two-phase learning framework for radio map estimation
by integrating radio propagation model and designing a conditional generative 
adversarial network (cGAN). We
first explore global information to extract the radio propagation patterns. 
We then focus on the local features to estimate the effect of shadowing on
radio maps in order to train and optimize the cGAN. Our experimental results 
demonstrate the efficacy of the proposed framework for radio map estimation
based on generative models from sparse observations in outdoor scenarios.
\end{abstract}


\section{Introduction}\label{intro}
\begin{figure*}[t]
	\centering
	\subfigure[Radio Map over Different Locations]{
		\label{ex1}
		\includegraphics[height=3.8cm]{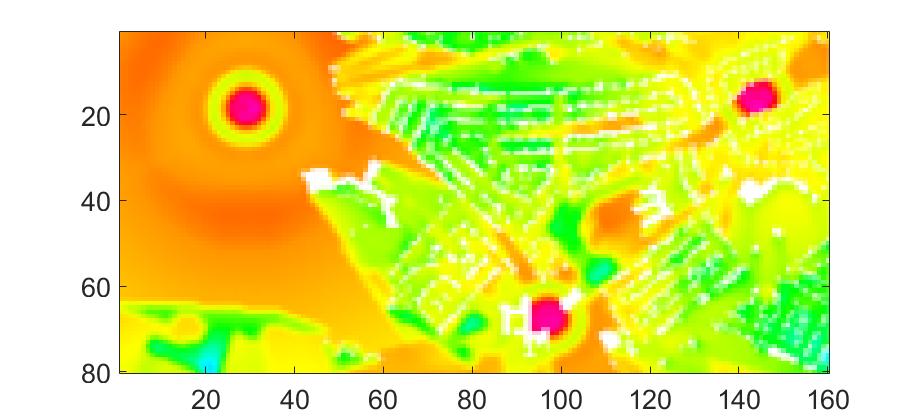}}
	\hspace{1cm}
	\subfigure[Sparse Samples Collected by Sensors]{
		\label{ex2}
		\includegraphics[height=3.8cm]{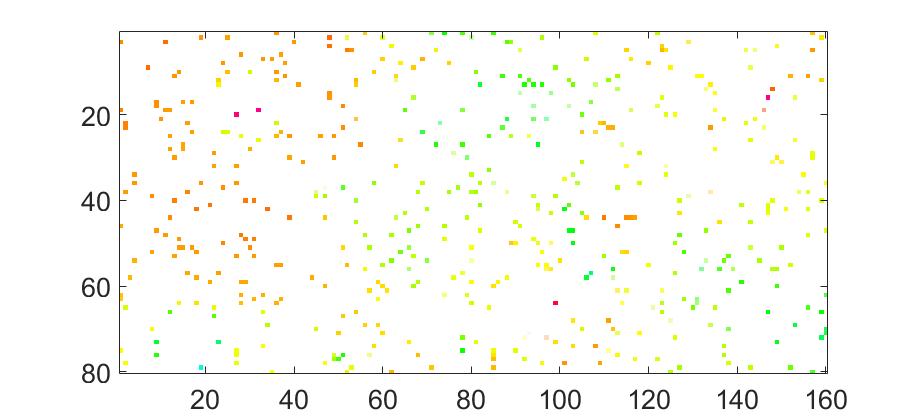}}
	\caption{Example of Radio Map: a) radio map generated from simulators with three transmitters; b) 1000 samples from the original radio map.}
	\label{ex_mln}
\end{figure*}

\IEEEPARstart{T}he rapid development of information technologies
has spawned many novel concepts and applications, such as the Internet of Things (IoT),
autonomous driving, and 
cellular networks. For these wireless services, efficient spectrum planning and network 
coverage analysis are critical to spectrum efficiency and user experience. Specifically, radio 
maps provide vital information on spatial distribution of radio frequency (RF) signal power, and 
facilitate resource allocation and network outage diagnosis \cite{a1}. 
In general, RF `radio map' refers to a geographical signal power spectrum density (PSD), 
formed by cumulative received RF signal power, as a function of spatial locations and frequencies,
and may also exhibit temporal variations \cite{a2}. Enriching useful information on propagation 
behavior and spectrum occupancy, radio map could provide more detailed information
on PSD distribution \cite{a3} spatially, shown as Fig. \ref{ex1}. 
However, RF signal powers can only be sparsely measured in practice. 
Thus, a practical challenge is the estimation of fine-resolution radio map 
from spatially sparse signal power samples collected by sensors or user devices as shown in 
Fig.~\ref{ex2}. Our goal here is to achieve more efficient and accurate estimation of radio maps 
from sparse observations.

Most current approaches to radiomap estimation (reconstruction) can be categorized into
either model-based or model-free. Model-based methods usually assume certain signal 
propagation models and express the received signal PSD as a combination from  
active transmitters \cite{a2}. For example, the authors of \cite{a4} introduced 
log-distance path loss model (LDPL) for Wi-Fi radio map reconstruction. 
In \cite{a5}, another model-based method proposed interpolation based on thin-plate splines kernels. 
On the other hand, model-free methods do not assume a specific model but explore more
neighborhood information. Typical examples include Radial Basis Function (RBF) \cite{a6} 
interpolation and inverse distance weighted (IDW) interpolation \cite{a7}. 
In addition to interpolation methods, there have also been a number of works that apply machine learning 
for radio map reconstruction. The authors of \cite{a8} presented a fast radio map estimation 
based on Unet \cite{k1}. Another team \cite{a9} proposed deep auto-encoders for 
estimating geographical PSD distribution. We will provide a more complete literature review
later in Section \ref{related}.

Despite many successes, existing solutions still exhibit certain limitations. First, unlike
the problem of astronomical image analysis \cite{a10} and 3D city map reconstruction \cite{a11}, 
ground truth radio map is more sensitive to the complex environment 
beyond mere radio propagation models. Failure to capture effects from shadowing or 
obstacles limits the efficacy of model-based approaches. On the other hand, 
the performance of model-free methods relies on the quality of observed samples. 
Generally, they assume a uniform distribution of measurements over the geometric space. 
However, in practical applications such as minimization of drive test (MDT) \cite{a12}, 
measurements are collected from user devices, which are typically non-uniformly distributed within
cell coverage. It is a challenge to process unevenly distributed observations efficiently.
Moreover, samples in different training sets do not necessarily have the same radio propagation models and parameters. In view of their respective strengths
and shortcomings, we should consider effective integration of model-based and model-free methods.

To address the aforementioned problems, we investigate radio propagation in the outdoor scenario, and propose a two-phase learning framework (RME-GAN) for radio map estimation based on conditional generative adversarial networks (cGAN) \cite{a31}. Since a radio map can be dynamically constructed from real-time measures and continuously updated \cite{a2}, we mainly focus on the step of
radio map estimation from sparse observations over geographical locations.
In phase 1, we integrate learning-based and model-based principles to 
extract global information while leveraging radio propagation model. In phase 2, we utilize 
different sampling strategies to capture local radiomap features and to explore the shadowing effect.
More specifically, we propose to capture the sharp features in the Fourier domain for detailed information, 
before resampling the measured signals geometrically to address the non-uniformity among
distributed observations. 
We summarize our contributions as follows:
\begin{itemize}
    \item We introduce a strategy to integrate well known radio propagation model into
    data-driven approaches to guide the learning machines for global radiomap information in the first phase of our proposed framework;
    \item We design different sampling strategies to capture  local radio information and shadowing effects to overcome the non-uniformity of distributed observations in the second phase of the proposed framework;
    \item We provide a comprehensive experimental comparison with state-of-the-art methods in terms of both reconstruction accuracy and fault diagnosis;
\end{itemize}
The experimental results demonstrate the ability of the proposed strategies to capture global model information and local shadowing features, as well as the power of the two-phase learning frameworks in radio map reconstruction.

We organize
the remainder of this work 
as follows. We first overview the related works of radio map reconstruction and generative adversarial network (GAN) models in Section \ref{related}. We then introduce our problem formulation in Section \ref{Pro}. In Section \ref{Fra}, we present the details of the proposed two-phase learning frameworks based on cGAN. Following the presentation of the experimental results in Section \ref{Exp}, we conclude our works in Section \ref{Con}.

\section{Related Work} \label{related}
Here we provide an overview of radio map estimation from
sparse samples based on either model-based or model-free approaches. 
We also introduce GAN models.
\subsection{Radio Map Estimation}
\subsubsection{Model-based Radio Map Estimation}
Model-based approaches usually leverage a certain signal propagation model. 
Specifically, a radio map function is modeled as a function of
frequency $f$ and a geographic location $\mathbf{c}$ by
\begin{equation}
    \mathbf{\Phi}(\mathbf{c},f)=\sum_{i=1}^{N_t}g_i(\mathbf{c},f)\Psi_i(f)
\end{equation}
where $N_t$ denotes the number of transmitters, $g_i$ denotes the channel power gain of
the $i$th transmitter, and $\Psi_i(f)$ is the PSD of the the $i$th 
transmission \cite{a2}. Here, $g_i$ can either be assumed as an explicit function with prior knowledge depending on specific tasks, or be represented in terms of sensor 
parameters. For example, the authors of \cite{a4} 
modeled a single narrowband WiFi transmitter with $g_i$ 
according to the log-distance path loss model (LDPL). In \cite{a5}, the authors modeled $g_i$ as a kernel expansion to estimate the power gain. Other model-based methods include parallel factor analysis (PARAFAC) \cite{a13} and fixed rank kriging \cite{a14}.

\subsubsection{Model-free Radio Map Reconstruction}
Different from model-based methods, model-free methods usually do not assume a specific propagation model but favor neighborhood information. In general, the model-free methods can be categorized as interpolation-based methods and learning-based methods as follows:
\begin{itemize}
    \item Interpolation-based Methods: Interpolation-based approaches express the PSD at a particular location as a combination of neighborhood measurements, i.e.,
    \begin{equation}
        \mathbf{\Phi}(\mathbf{c},f)=\sum_{i=1}^{N_s}w_i(\mathbf{c},f)q_i(f),
    \end{equation}
    where $q_i$ denotes the observation from the $i$th observation and $w_i$ is the 
    combination weight \cite{a2}.
    One typical example is inverse distance weighted (IDW) interpolation \cite{a7}, where the power gain is inversely proportional to the distance between transmitters and receivers. Besides linear interpolation, an effective alternative is radio basis function (RBF) interpolation \cite{a6}, where different kernels, such as Gaussian, multiquadrics, or spline, can be applied. Another classic interpolation approach is ordinary Kriging methods \cite{a15}, where the PSD is optimized by minimizing the variance of estimation error. Beyond these traditional methods, recent works of radio map interpolation studied integration with novel data analysis concepts, such as graph signal processing \cite{a16}, crowdsourcing methods \cite{a17}, image inpainting \cite{a18,a19} and tensor completion \cite{a20}.
    
    \item Learning-based Methods: Recent applications and success of machine learning in wireless networks and IoT intelligence have stimulated more interests in learning-based radio map estimation as another promising direction. Different from explicit interpolation, learning-based methods focus on finding a direct mapping from an input geometric map to its output PSD measurement. The functional relationship between input and output is usually captured
    by a neural network. For example, Unet is introduced in \cite{a8} for predicting the pathloss. Similarly, a deep auto-encoder is another design for the deep neural network
    mapping \cite{a9}. Other learning frameworks for radio map estimation include transfer learning \cite{a21}, generative adversarial network \cite{a22,a23, a24}, deep Gaussian process \cite{a25}, reinforcement learning \cite{a26}, and deep neural networks \cite{a27,a28,a29}.
\end{itemize}
Interested readers could refer to \cite{k2} for a more thorough overview of data-driven radio map reconstruction.

\subsection{Generative Adversarial Networks}
The idea of 
Generative Adversarial Networks (GAN) is to study a
collection of training examples to acquire sufficient information on sample
distribution in order to generate data samples with similar distribution \cite{a30}. In GAN, two neural networks interact in the form of a zero-sum game, in which the generator $G$ aims to generate new data with the same statistics as the training dataset while the discriminator $D$ focuses on the identification of the true and fake data samples. 
Suppose the generator learns the distribution $p_g$ over data $\bf{x}$ by mapping a prior noise distribution $p_z(\bf{z})$ to data space. The GAN model can be trained via
\begin{align}
    \min_G\max_D \quad V(D,G)= \mathbb{E}_{x\sim p_{\rm data}(\bf{x})}[\log D(\bf{x})] +\nonumber\\
    \mathbb{E}_{z\sim p_{z}(\bf{z})}[\log (1-D(G(\bf{z})))],
\end{align}
where $\mathbb{E}[\cdot]$ refers to the expectation operation.

\begin{figure*}[t]
	\centering
	\includegraphics[width=5in]{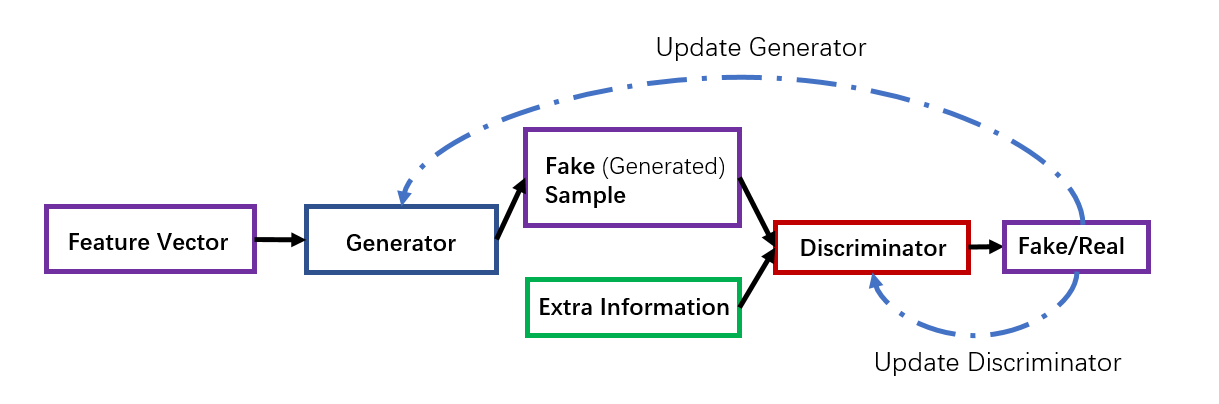}
	\caption{Example of a Customized cGAN.}
	\label{flat}
\end{figure*}

With the development of machine learning, GAN can be extended to a conditional model (cGAN) if the generator and discriminator are conditioned on some extra information $\mathbf{I}$ \cite{a31}. Then, the objective function could be set as 
\begin{align}
    \min_G\max_D \quad V(D,G)= \mathbb{E}_{x\sim p_{data}(\bf{x})}[\log D(\bf{x|I})] +\nonumber\\
    \mathbb{E}_{z\sim p_{z}(\bf{z})}[\log (1-D(G(\bf{z|I})))].
\end{align}

Note that, previous works have shown that cGAN can be customized to learn the desired properties by mixing the traditional loss function with cGAN objectives \cite{a32} and utilizing the corresponding feature vectors as input. See the illustration in Fig. \ref{flat}. For example, to generate a painted image from sketch, one can input the sketch to the generator and formulate the training process as a one-to-one mapping.
No random noise vector should be involved for such mapping problem. 
The customized cGAN frameworks have achieved successes in applications, including image painting \cite{a33}, image-to-image translation \cite{a34}, and image manipulation \cite{a35}.

\section{Problem Formulation} \label{Pro}
Now, we present the problem formulation in this work. In general, we focus on the radio map reconstruction from sparse observations in the outdoor scenario.

Our model considers a set of regions with sparse observations of network coverage, similar to
Fig. \ref{ex2}. Each region is conformed to a size of $256\times 256$ grid and contains $N_t$ transmitters with known positions. Let $\mathbf{P_t}_i\in\mathbb{R}^{Nt\times 2}$ denote the 2-dimensional (2D) coordinates of transmitters for the $i$th region.
Region $i$ is also attributed with their corresponding urban maps $\mathbf{m}_i\in\mathbb{R}^{256\times 256}$ with building information.
Instead of knowing the fine-resolution radio map of the whole space, each region only observes a set of $K_i$ sparse samples $(\mathbf{x}_i,\mathbf{c}_i)$ collected from sensors, where $K_i\leq 256\times 256$, $\mathbf{x}_i\in\mathbb{R}_+^{K_i}$ are the non-negative observed PSD, and $\mathbf{c}_i\in\mathbb{R}^{K_i\times 2}$ are the 2D coordinates of the corresponding observations. Thus, each region $i$ is characterized by 
\begin{equation}
    \mathcal{F}_i=\{(\mathbf{x}_i,\mathbf{c}_i), \mathbf{m}_i,\mathbf{P_t}_i\}.
\end{equation}
%\textcolor{red}{Should be focus on real positive values in the radio map?  If so, it is
%better to use $\mathbb{R}_+$.}

Suppose that some regions have the ground truth of the fine-resolution radio map $\mathbf{y}$. Treating these regions as training regions, our goal is to train a learning machine $g(\cdot)$ to estimate the fine-resolution radio map for all regions, i.e., 
\begin{equation}\label{form}
    \tilde{\mathbf{y}}_i=g(\mathcal{F}_i)\in\mathbb{R}_+^{256\times 256}.
\end{equation}

Unlike existing learning-based radio map reconstruction, we do not 
apply strict 
constraints on the distribution of samples and propagation model parameters.
The following conditions are relaxed:

\begin{itemize}
    \item The sparse observations $\mathbf{x}_i$ can be either uniformly or unevenly distributed over the whole region;
    \item Different regions may have different number $K_i$ of observations;
    \item The transmitters in each region may have different radio propagation parameters.
\end{itemize}

To deal with such non-uniformly distributed observations and the unbalanced data samples over different training regions, we propose a two-phase learning framework for radiomap estimation (RME) based on GAN, known as RME-GAN. Our goal is to train an efficient generator $g(\cdot)$ to estimate the full radio map as described in Eq. (\ref{form}).

\section{Radio Map Reconstruction via RME-GAN} \label{Fra}
\begin{figure*}[t]
	\centering
	\subfigure[Structure of Generator]{
		\label{ex3}
		\includegraphics[height=5.5cm]{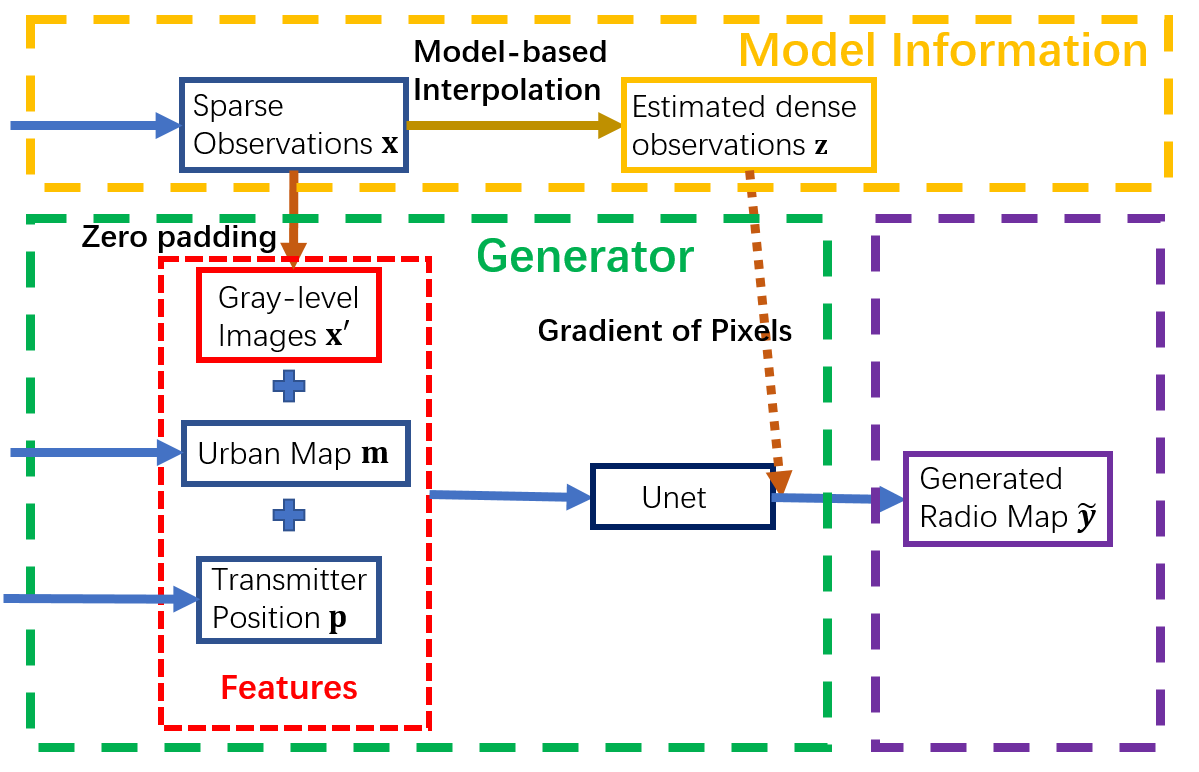}}
	\hspace{0.5cm}
	\subfigure[Structure of Discriminator]{
		\label{ex4}
		\includegraphics[height=5.5cm]{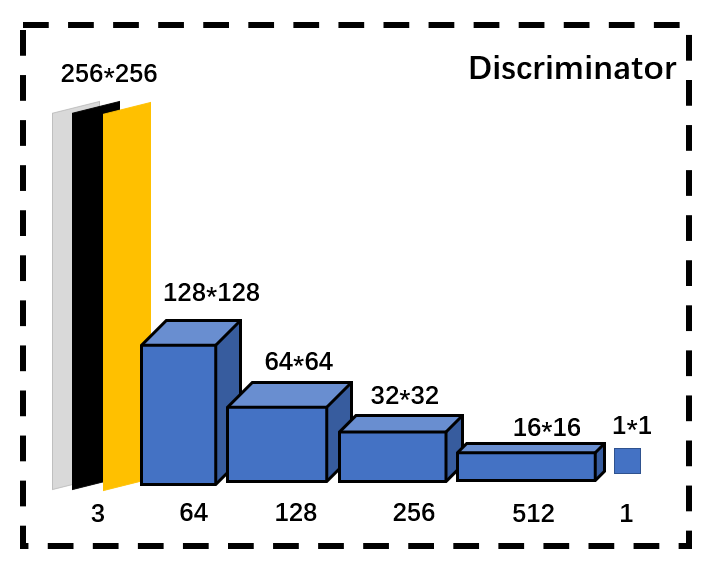}}
	\caption{Structure of RME-GAN: a) The generator takes input $\mathcal{F}_i$ 
 and consider the model information during training, where two phases of loss functions are applied to train the generator; b) The discriminator consists of {4} layers and generates a one-dimensional probability output. {The discriminator input consists of 3 channels: 
 the generated/true radio map (marked as orange), and the prior information as two-channel one-hot encodings (marked as grey and black). } }
 %The input of discriminator are the prior information (marked as red) and the generated/true radio map (marked as orange).}
	\label{ex_gan}
\end{figure*}

In this section, we present the design of the proposed RME-GAN, with the overall structure shown as Fig. \ref{ex_gan}.
We first introduce the basic sketch of RME-GAN in Section \ref{skk},
before describing the two-phase customization of features and loss functions in Section \ref{two_phase} and Section \ref{loss}, respectively.

\begin{table*}[t]
\centering
\caption{Structure of Unet}
\begin{tabular}{|l|l|l|l|l|l|l|l|l|l|l|}
\hline
\rowcolor[HTML]{C0C0C0} 
Layer      & Input   & 1       & 2       & 3       & 4     & 5     & 6     & 7      & 8    & 9      \\ \hline
Resolution & 256     & 256     & 128     & 64      & 64    & 32    & 32    & 16     & 8    & 4      \\ \hline
Channel    & 3       & 6       & 40      & 50      & 60    & 100   & 100   & 150    & 300  & 500    \\ \hline
Filter     & 3       & 5       & 5       & 5       & 5     & 3     & 5     & 5      & 5    & 4      \\ \hline
\rowcolor[HTML]{C0C0C0} 
Layer      & 10      & 11      & 12      & 13      & 14    & 15    & 16    & 17     & 18   & Output \\ \hline
Resolution & 8       & 16      & 32      & 32      & 64    & 64    & 128   & 256    & 256  & 256    \\ \hline
Channel    & 300+300 & 150+150 & 100+100 & 100+100 & 60+60 & 50+50 & 40+40 & 20+6+3 & 20+3 & 1      \\ \hline
Filter     & 4       & 4       & 3       & 6       & 5     & 6     & 6     & 5      & 5    & -      \\ \hline
\end{tabular}
\label{unet}
\end{table*}

\subsection{Sketch of RME-GAN}\label{skk}
RME-GAN learns a mapping function $g(\cdot)$ from the region features $\mathcal{F}_i$ to its output estimated full radio map $\tilde{\mathbf{y}}_i$. Since it is also a one-to-one mapping problem, we could design RME-GAN based on the general cGAN structure similar to Fig. \ref{flat}. More specifically, we utilize an 18-layer deep Unet \cite{a8} structure for the sketch of the generator, with parameters shown in Table. \ref{unet}.

To reformat the input features in $\mathcal{F}_i$, we first interpolate the sparse observations $(\mathbf{x}_i,\mathbf{c}_i)$ to the same size as region size via zero padding, i.e., $\mathbf{x}_i'\in\mathbb{R}_+^{256\times 256}$. Since we consider a convolutional neural network (CNN) structure for Unet, the interpolated zero would not influence other features. We then encode the transmitter locations $\mathbf{P_t}_i$ also in a $256\times 256$ matrix $\mathbf{p}_i$, where the transmitter positions in $\mathbf{p}_i$ are marked as $1$ and other positions are $0$. Features $\mathcal{F}_i$ are encoded to three channels of features with size $256\times 256$, i.e., $\mathbf{f}_i=\{\mathbf{x}_i',\mathbf{m}_i,\mathbf{p}_i\}$.

With the encoded features, the basic objective function of RME-GAN can be designed by
\begin{align}
    \min_G\max_D \quad V(D,G)= \mathbb{E}_{y,I\sim p_{\rm data}(\bf{y},\bf{I})}[\log D(\bf{y,I})] +\nonumber\\
    \mathbb{E}_{f\sim p_{\rm data}(\bf{f})}[\log (1-D(G(\bf{f}),\mathbf{I}))],
\end{align}
where $p_{\rm data}(\bf{y,I})$ is the joint distribution with the prior knowledge $\mathbf{I}$. More specifically, $\mathbf{I}$ can be designed as a function of the region features, i.e., $\mathbf{I}=f(\mathcal{F}_i)$ or some determined labels, such as fake or real \cite{a33}. Here, we set $\mathbf{I}\in\mathbb{R}^{256\times 256\times 2}$ {to represent the one-hot encodings in two-channel to give prior information to the discriminator as the input is a generated or real radio map.}

%all ones if the input of discriminator is generated radio map; otherwise, we set it with all zeros. (\textcolor{red}{check 1 or 0})

We can split the min-max optimization problem to sub-problems, and train the generator $G$ and discriminator $D$ iteratively. Specifically, the loss function to optimize $D$ is
\begin{equation}
    L_D=-V(D,G),
\end{equation} 
and the general loss function to optimize $G$ is 
\begin{equation}\label{gan_obj}
    L_G=\mathbb{E}_{x\sim p_{\rm data}(\bf{x})}[\log (1-D(G(\bf{x}),\mathbf{I}))].
\end{equation}

The overall sketch of RME-GAN is illusrated in Fig. \ref{ex_gan}.
Note that the cGAN can be customized by adding additional loss terms to capture 
desired properties. In this work, we consider a two-phase loss function design for
Eq.~(\ref{gan_obj}) to capture the global and local information, respectively. More details will be discussed in Section \ref{two_phase} and \ref{loss}.

\subsection{Two-phase Customization in RME-GAN} \label{two_phase}
Earlier, we have identified two challenges in radio map estimation: 
1) how to efficiently integrate radio propagation model with learning machines to capture 
global information; and 2) how to estimate the details from biased and non-uniformly distributed observations. The first challenge lies in the estimation of global model information while the second challenge deals with shadowing effect and obstacle impact. To handle these challenges, we utilize different sampling strategies to extract global patterns and local features, 
respectively, as follows.
\subsubsection{Upsampling for global information}
In general, radio propagation should follow certain pathloss models, such as LDPL \cite{a4}. Thus, to model global pathloss, we first upsample the sparse observations $(\mathbf{x}_i,\mathbf{c}_i)$ to a radio map template $\mathbf{z}\in\mathbb{R}_+^{256\times 256}$ via model-based interpolation (MBI). Suppose that a system has $N_t$ transmitter with positions $(a_k,b_k)$, $1\leq k \leq N_t$. In MBI, the signal strength (in dB) at position $(a,b)$ can be modeled as
\begin{equation}
    P(a,b)=\sum_{k=1}^{N_t} (\alpha_k-10\theta_k \log_{10}(\sqrt{(a_k-a)^2+(b_k-b)^2})),
\end{equation}
where $\alpha_k$ is the power from $k$th transmitter at the reference distance. Then, we optimize the paramters of the LDPL model from the observations by
\begin{equation}
    \min_{\alpha_k,\theta_k} \sum_{j=1}^{K_i} (\mathbf{x}(j)-P(\mathbf{c}(j)))^2
\end{equation}
where $\mathbf{x}(j)$ is the $j$th observation with position $\mathbf{c}(j)$. By apply the LDPL model with estimated parameters, we can upsample the sparse observations $\mathbf{x}_i$ as $\mathbf{z}_i\in\mathbb{R}_+^{256\times 256}$.

Although MBI may be not able to capture full effect of obstacles and shadowing, 
upsampled template can still capture the global propagation pattern together with MBI model parameters. Consideration of upsampled features during the training process can provide 
a general guideline on radio map reconstruction. We introduce how to use $\mathbf{z}_i$ to design loss functions of Phase 1 in Section \ref{loss}.

\subsubsection{Downsampling for detailed features}
After estimating a template of reconstructed radio maps, we downsample the sparse observations in the second phase to correct errors in the details. We consider two downsampling strategies:
\begin{itemize}
    \item Geometry-wise downsampling: As discussed in Section \ref{Pro}, sparse observations may be non-uniformly distributed over the whole space. To deal with unevenly distributed samples, we apply geometric information to downsample the observations. To balance the sample numbers in different locations, we first build superpixels \cite{a36} based on urban map or interpolated template to split the entire region into $N_s$ pieces. In each piece, we only pick one observation of peak energy. Then, we have a downsampled version of observations $\mathbf{x}\in\mathbb{R}_+^{K}$ as $\mathbf{x_s}\in\mathbb{R}_+^{N_{s}}$. In addition to superpixel-based space cut, other methods such as Voronoi diagram \cite{a4} or spectral clustering \cite{a37} can be also applied.
    \item Frequency-wise downsampling: In digital signal processing, 
    details and outliers are related to high frequencies in Fourier space. Thus, we could also transform the estimated radio map $\tilde{\mathbf{y}}$ to Fourier space as $\hat{\mathbf{y}}$. Then, we can select the first $N_f$ Fourier coefficients in the high frequency part as additional information in loss functions of the second phase, i.e., $\hat{\mathbf{y}}_f\in\mathbb{R}^{N_f}$. 
\end{itemize}

With the extracted features under different sampling strategies, we
propose a two-phase design on the loss functions for the training of the generator, i.e.,
\begin{equation}
    L_G'=\left\{
    \begin{aligned}
    &L_G+L_{Global} \quad& \text{if}\quad \epsilon<\tau\\
    &L_G+L_{Local} \quad & \text{if}\quad \epsilon\geq\tau
    \end{aligned}
    \right.,
\end{equation}
where $L_{Global}$ consist of loss functions favoring global model information and $L_{Local}$ are the loss functions focusing on the local details. Here, a decision variable $\epsilon$ can be designed according to certain criteria, i.e., length of epochs or validation accuracy. For convenience, we keep a validation data set to evaluate the validation accuracy at each epoch to determine the phase change point. The specific design of loss functions will be introduced 
next in Section \ref{loss}.

\subsection{Loss Functions}\label{loss}
After introducing the overall frameworks, we now present the design of loss functions.
\subsubsection{Phase 1 - Global Model Prediction}
We first introduce the design of loss function in the first phase.
\begin{itemize}
    \item \textit{Baseline loss function}: The first term is the mean squared error (MSE) between the generated radio map and the ground truth $\mathbf{y}$:
    \begin{equation}
        L_{\rm MSE}=\frac{1}{256\times 256}\cdot ||\mathbf{y}-\tilde{\mathbf{y}}||_F^2,
    \end{equation}
    where $||\cdot||_F$ is the Frobenius norm and $\tilde{\mathbf{y}}=g(\mathbf{f})$.
We also consider a second term to be the total variation which describes the smoothness of the generated radio map \cite{a33}. More specifically, we apply 
the total variation of images 
    \begin{equation}
        L_{\rm TV}=\sum_{i,j\in\mathcal{N}}||\tilde{\mathbf{y}}(i)-\tilde{\mathbf{y}}(j)||_2^2,
    \end{equation}
    where $\mathcal{N}$ indicates pixel neighborhood and $\tilde{\mathbf{y}}(i)$ is the $i$th element of the estimated radio map $\tilde{\mathbf{y}}$.
    \item \textit{Loss function of global information}: In addition to the baseline loss functions, we also utilize upsampled radio map $\mathbf{z}$ from MBI to help the learning of global radio propagation patterns. Instead of directly evaluating the difference between $\tilde{\mathbf{y}}$ and $\mathbf{z}$, we focus on gradient patterns. Such gradient patterns feature  smoothness of radio propagation from transmitters to the surroundings. Since the first phase aims to learn the radio propagation model, MBI can help RME-GAN to learn the propagation model and speed up training. To obtain the gradient-based loss function, we first calculate the gradient of each pixel in the radio map for up, down, right and left, i.e.,
    \begin{equation}
        \mathcal{G}(\tilde{\mathbf{y}}(i))=[\mathcal{G}_{\rm up}, \mathcal{G}_{\rm down}, \mathcal{G}_{\rm right}, \mathcal{G}_{\rm left}]\in\mathbb{R}^{4}.
    \end{equation}
    Then, we calculate the cosine similar between $\mathcal{G}(\tilde{\mathbf{y}}(i))$ and $\mathcal{G}({\mathbf{z}}(i))$. Then, the loss function of global information is defined as
    \begin{equation}
        L_{\rm Gradient}= \sum_{i=1}^{256\times 256} CS(\mathcal{G}(\tilde{\mathbf{y}}(i)), \mathcal{G}({\mathbf{z}}(i))),
    \end{equation}
    where $CS(\cdot)$ refers to the cosine similarity \cite{a38}.
\end{itemize}

\subsubsection{Phase 2 - Local Shadowing Estimation}
Next, we introduce the loss functions for local features in the second phase.
\begin{itemize}
    \item \textit{Baseline loss function}: Similar to the first phase, the generated radio map should be smooth over the entire space. Thus, we also use total variation $L_{\rm TV}$ as one of the baseline loss functions. In addition, the $L_{\rm MSE}$ is also used here to control the overall accuracy. However, to emphasize more details depending on the downsampled observations, we assign a much smaller weight on $L_{\rm MSE}$ in the second phase compared to the first phase.
    \item \textit{Geometric loss function}: To mitigate the effect of non-uniformly distributed observations, we favor downsampled observations among all observations. Since the sparse observations are also true power gain in the radio map, we define a geometric loss function as the MSE between downsampled $\mathbf{x}_s=\mathbf{y}_s$ and the generated PSD at the same positions, denoted by $\tilde{\mathbf{y}}_s$, i.e.,
    \begin{equation}
        L_{\rm Geo}=MSE(\mathbf{x}_s,\tilde{\mathbf{y}}_s).
    \end{equation}
    \item \textit{Frequency domain loss function}: As discussed before in Section \ref{two_phase}, high frequency coefficients in the Fourier space can capture the details or outliers in a radio map. To capture such information, we also calculate the MSE between $\hat{\mathbf{y}}_f$ and $\hat{\tilde{\mathbf{y}}}_f$ denoted by
    \begin{equation}
        L_{\rm HPF}=MSE(\hat{\mathbf{y}}_f, \hat{\tilde{\mathbf{y}}}_f),
    \end{equation}
    where $\hat{\mathbf{y}}_f$ includes the first top $N_f$ high frequency coefficients in the Fourier domain.
    \item \textit{Multi-scale structural similarity index}: In image processing, multi-scale structural similarity index (MS-SSIM) is a useful term to evaluate the image quality based on human visual system \cite{a39}. Since details in radio map can be viewed as a fine-resolution image while the model-based template can be viewed as a coarse-resolution image, we could also consider the MS-SSIM index to improve the quality of generated radio map in the second phase. Let $\mathbf{u}={u_i|i=1,2,\cdots, N}$ and $\mathbf{v}={v_i|i=1,2,\cdots, N}$ be two discrete non-negative signals, e.g., two image patches from the same position or two radio map residing on the same region. The luminance, contrast and structure comparison measurements can be calculated as follows:
    \begin{equation}
        l(\mathbf{u},\mathbf{v})=\frac{2\mu_u\mu_v+C_1}{\mu_u^2+\mu_v^2+C_1},
    \end{equation}
    \begin{equation}
        c(\mathbf{u},\mathbf{v})=\frac{2\sigma_u\sigma_v+C_2}{\sigma_u^2+\sigma_v^2+C_2},
    \end{equation}
    \begin{equation}
        s(\mathbf{u},\mathbf{v})=\frac{2\sigma_{uv}+C_3}{\sigma_u\sigma_v+C_3},
    \end{equation}
    where $\mu$ represents mean, $\sigma$ is the variance, $C_i$ are small constants introduced in \cite{a40}.
    To calculate MS-SSIM, the system iteratively applies a low-pass filter and downsamples the filtered image by a factor of 2 on reference and distorted image signals
    as inputs. Suppose that $M$ is the highest scale. The overall MS-SSIM is calculated by
    \begin{equation}
       {\rm SSIM}(\mathbf{u},\mathbf{v})=[l_M(\mathbf{u},\mathbf{v})]^{\alpha_M}\cdot\prod_{j=1}^M[c_j(\mathbf{u},\mathbf{v})]^{\beta_j}[s_j(\mathbf{u},\mathbf{v})]^{\gamma_j}.
    \end{equation}
    Here, ${\rm SSIM}(\mathbf{u},\mathbf{v})\leq1$, and ${\rm SSIM}(\mathbf{u},\mathbf{v})=1$ \textit{iff} $\mathbf{u}=\mathbf{v}$.
    The MS-SSIM loss function is defined by
    \begin{equation}
        L_{\rm SSIM}=1-{\rm SSIM}(\mathbf{y},\tilde{\mathbf{y}}).
    \end{equation}
    More properties of MS-SSIM can be found in \cite{a39}.
\end{itemize}

\subsection{Summary of Two-Phase Training Strategy}
Combining all the loss terms aforementioned, the final loss function to train the generator is represented as 
\begin{align}
    &L_G'=\left\{
    \begin{aligned}
    &\lambda_1 L_G+\lambda_2 L_{\rm MSE}+\lambda_3 L_{\rm TV} + \lambda_4 L_{\rm Gradient}, & \epsilon<\tau\\
    &\lambda_1L_G+\lambda_2 L_{\rm MSE}+\lambda_3 L_{\rm TV}+\\ &\lambda_5L_{\rm SSIM}+\lambda_6 L_{\rm Geo}+\lambda_7 L_{\rm HPF} &  \epsilon\geq\tau
    \end{aligned}
    \right. \nonumber
\end{align}
where $\lambda_i$ are the weights to control each loss term.
In the first phase, we aims to apply model-based interpolation to help RME-GAN learn a template to capture the radio propagation model. After enough epochs, we favor more on the local information to correct errors in the details on the template by learning the shadowing effect and obstacle impact. Although some baseline loss functions are shared in both phases, the weights can be different depending on the specific tasks.

\section{Experimental Results} \label{Exp}
In this section, we present the experimental results of the proposed methods.

\subsection{Experiment Setup} 
\begin{figure}[t]
	\centering
	\subfigure[Example of Urban Images.]{
		\label{ex5}
		\includegraphics[height=4cm]{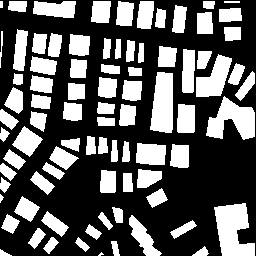}}
	\hspace{0.2cm}
	\subfigure[Example of Radio Map]{
		\label{ex6}
		\includegraphics[height=4cm]{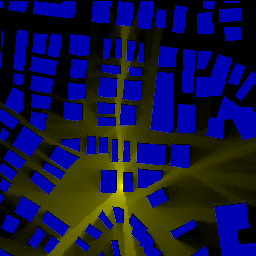}}
	\caption{Examples of RadioMapSeer Dataset: a) Urban maps with buildings as white and background as black; b) radio map with pathloss as yellow.}
	\label{ex_rm}
\end{figure}

\begin{figure}[t]
	\centering
	\subfigure[Ground Truth]{
		\label{gt}
		\includegraphics[height=3cm,width=3cm]{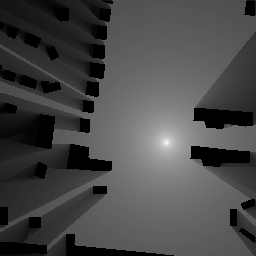}}
	\hspace{0.2cm}
	\subfigure[Three Selected Regions]{
		\label{cut}
		\includegraphics[height=3cm,width=3.1cm]{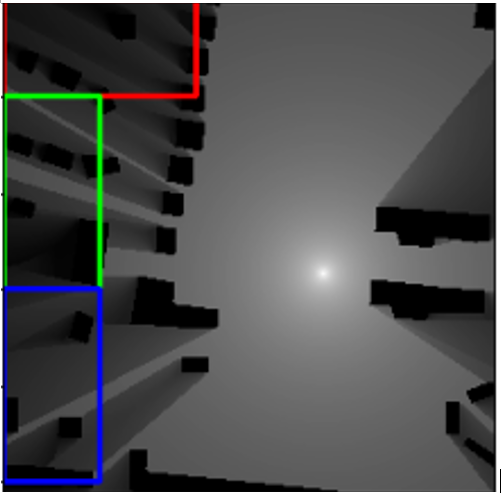}}
	\caption{One Example of Radio Map}
	\label{ex_3}
\end{figure}

\subsubsection{Dataset} In this work, we test the performance of the proposed method in the RadioMapSeer dataset\footnote{\url{https://radiomapseer.github.io/}} \cite{a8}. The RadioMapSeer dataset consists of 700 maps with 80 transmitter locations per map. The city maps are taken from OpenStreetMap \cite{a41} covering metropolitans such as Ankara, Berlin, Glasgow, Ljubljana, London and Tel Aviv. The heights of the transmitters, receivers and buildings are set to 1.5m, 1.5m, and 25m, respectively, in this dataset. The transmitter power equals 23dBM and the carrier
frequency is 5.9GHz. The radio maps are generated using the software \textit{WinProp} \cite{a42}, and are saved in a 2D grid of $256\times 256 m^2$. The radio map has a 1-meter resolution. More specifically, we test on the higher accuracy simulations in the RadioMapSeer datasets.
%where two transmitters are considered in each map, i.e., $N_t=2$.
%with 300 receiver locations}. 
An example of the RadioMapSeer dataset is shown in Fig. \ref{ex_rm}, where building segmentation are consider as the urban map in our framework. The sparse observations are sampled from each radio map for training and testing.

\begin{figure*}[t]
	\centering
	\subfigure[NMSE]{
		\label{nmse}
		\includegraphics[height=7cm,width=8cm]{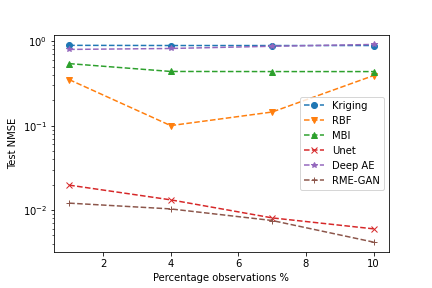}}
	\hfill
	\subfigure[RMSE]{
		\label{rmse}
		\includegraphics[height=7cm,width=8cm]{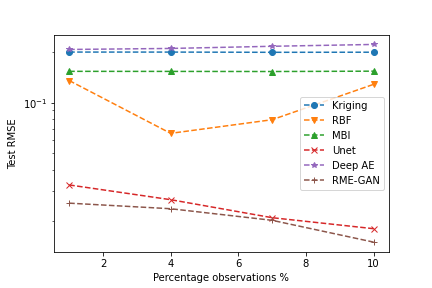}}
	\caption{Performance of Different Methods over Different Ratio of Observations.}
	\label{ratio}
\end{figure*}

\subsubsection{Data Splitting}
As mentioned in Section \ref{Pro}, the sparse observations can be either uniformly or non-uniformly distributed over the whole space, and different regions may have different number of observations. Thus, we sample the sparse observations under three different scenarios:
\begin{itemize}
    \item Setup 1 (Uniformly distributed): We uniformly sample $1\%$ of sparse observations from the whole radio map.
    \item Setup 2 (Unbalanced sample distribution among regions): In this setup, we sample the radio map at a random ratio from $1\%\sim 10\%$ in each region, and the sparse observations are uniformly distributed.
    \item Setup 3: (Non-uniformly distributed) We divide the radio map into two sides randomly. Then, we sample $1\%$ of observations from one side and $10\%$ from the other side.
 \end{itemize}
 We compare the proposed RME-GAN with other methods in all the three scenarios to demonstrate its efficacy. The whole dataset consists of 700 regions. For each experiment scenario, we split the dataset with 500 regions as training data, 100 regions for testing, and 100 regions for validation.

 \begin{table}[t]
\scriptsize
\centering
\caption{Performance in Different Setups}
\begin{tabular}{|l|ll|ll|ll|}
\hline
          & \multicolumn{2}{l|}{Setup 1}                           & \multicolumn{2}{l|}{Setup 2}                           & \multicolumn{2}{l|}{Setup 3}                           \\ \hline
Model     & \multicolumn{1}{l|}{NMSE}            & RMSE            & \multicolumn{1}{l|}{NMSE}            & RMSE            & \multicolumn{1}{l|}{NMSE}            & RMSE            \\ \hline
Kriging   & \multicolumn{1}{l|}{0.8863}          & 0.1947          & \multicolumn{1}{l|}{0.8987}          & 0.1962          & \multicolumn{1}{l|}{0.9118}          & 0.1971          \\ \hline
RBF       & \multicolumn{1}{l|}{0.3532}          & 0.1343          & \multicolumn{1}{l|}{0.1830}          & 0.0884          & \multicolumn{1}{l|}{0.3215}          & 0.1295          \\ \hline
MBI       & \multicolumn{1}{l|}{0.8837}          & 0.1973          & \multicolumn{1}{l|}{0.5158}          & 0.1499          & \multicolumn{1}{l|}{0.5174}          & 0.1500          \\ \hline
cGAN      & \multicolumn{1}{l|}{0.1559}          & 0.0904          & \multicolumn{1}{l|}{0.15077}                & 0.0888                & \multicolumn{1}{l|}{0.1506}                &     0.0889            \\ \hline
AE        & \multicolumn{1}{l|}{0.1823}          & 0.0978          & \multicolumn{1}{l|}{0.2885}                & 0.1238                & \multicolumn{1}{l|}{0.2783}                &     0.1217            \\ \hline
Deep AE   & \multicolumn{1}{l|}{0.1898}          & 0.0998          & \multicolumn{1}{l|}{0.3152}                &  0.1295               & \multicolumn{1}{l|}{0.3058}                &        0.1274         \\ \hline
Unet      & \multicolumn{1}{l|}{0.0093}          & 0.0220          & \multicolumn{1}{l|}{0.0053}                &    0.0166             & \multicolumn{1}{l|}{0.0050}                &    0.0161             \\ \hline
RadioUnet & \multicolumn{1}{l|}{0.0052}          & 0.0164          & \multicolumn{1}{l|}{0.0042}          & 0.0148          & \multicolumn{1}{l|}{0.0046}          & 0.0155          \\ \hline
RME-GAN    & \multicolumn{1}{l|}{\textbf{0.0043}} & \textbf{0.0151} & \multicolumn{1}{l|}{\textbf{0.0036}} & \textbf{0.0130} & \multicolumn{1}{l|}{\textbf{0.0038}} & \textbf{0.0140} \\ \hline
\end{tabular}
\label{oa}
\end{table}

\begin{figure*}[t]
	\centering
	\subfigure[RME-GAN]{
		\label{v1}
		\includegraphics[height=3.5cm]{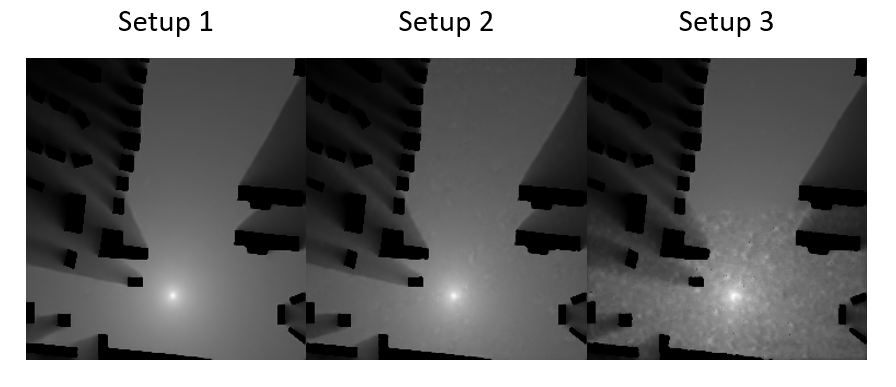}}
	\hspace{0.5cm}
	\subfigure[RadioUnet]{
		\label{v2}
		\includegraphics[height=3.5cm]{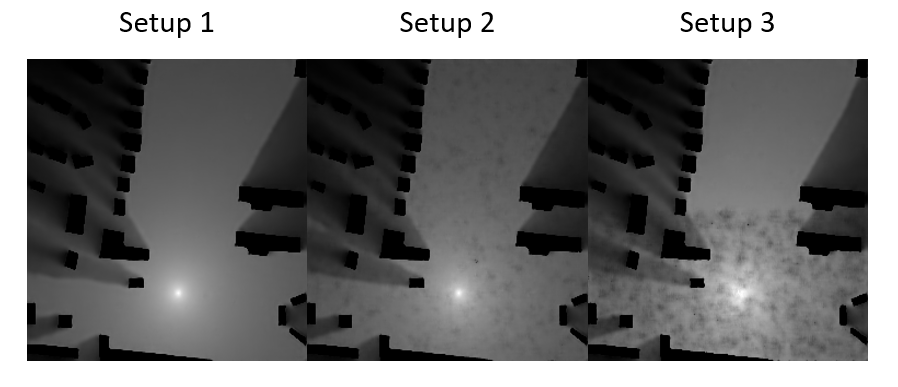}}
  \hspace{0.5cm}
	\subfigure[MBI]{
		\label{v3}
		\includegraphics[height=3.5cm]{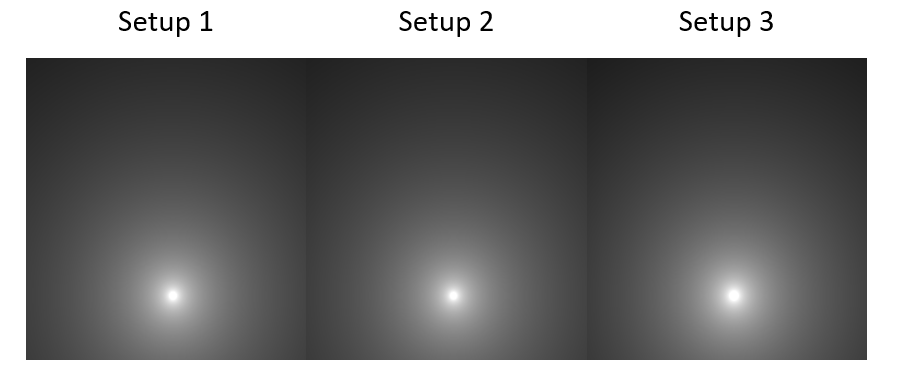}}
  \hspace{0.5cm}
		\subfigure[Kriging]{
		\label{v4}
		\includegraphics[height=3.5cm]{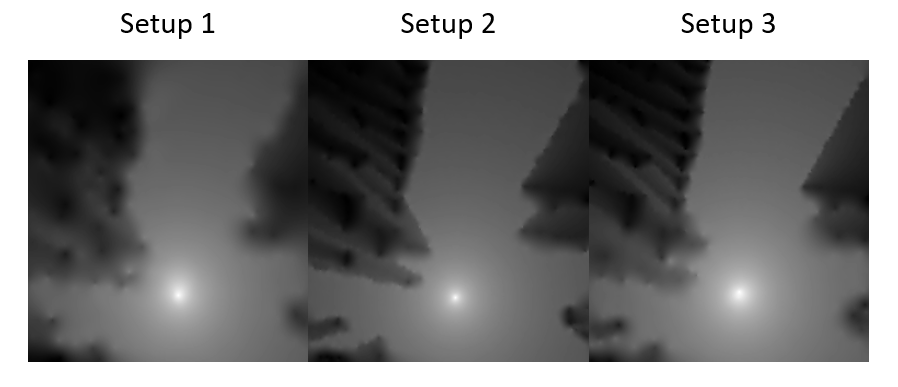}}
		\subfigure[RBF]{
		\label{v5}
		\includegraphics[height=3.5cm]{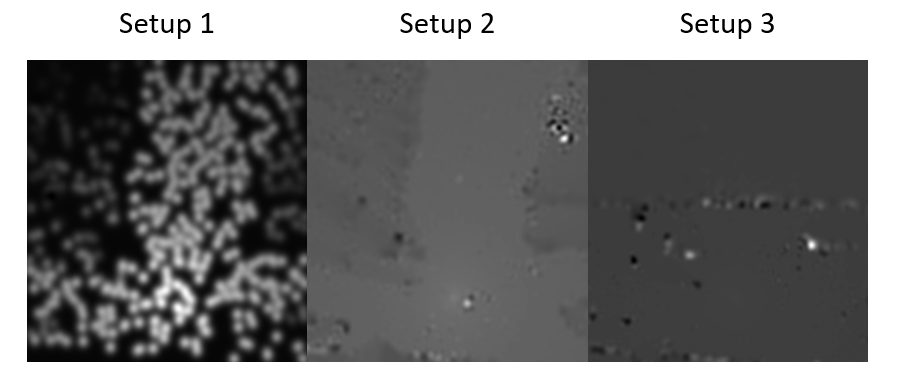}}
  \hspace{0.5cm}
		\subfigure[AE]{
		\label{v6}
		\includegraphics[height=3.5cm]{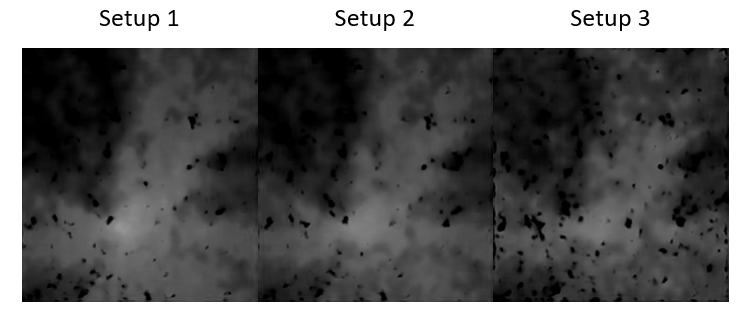}}
	\caption{Visualization Results of Different Methods }
	\label{VISUAL}
\end{figure*}

\subsubsection{Parameters of Learning Models}
We utilize an 18-layer Unet for the generator with parameters shown in Table. \ref{unet}, and a 4-layer discriminator shown as Fig. \ref{ex4}. In the first phase, the weights of each loss term are set as $\lambda_2=10$, $\lambda_3=0.01$, $\lambda_4=1$. In the second phase, we set the weights as $\lambda_1=10$, $\lambda_2=1$, $\lambda_3=0.001$, $\lambda_4=0.0001$, $\lambda_5=0.0001$ and $\lambda_6=0.0001$. Note that, although both phases contains the $L_{\rm MSE}$, $\lambda_2$ is set to be smaller in the second phase since we favor the downsampled radio map for detail correction.

\subsection{Overall Accuracy}
We first present the results in different setups. To evaluate the performance of the proposed RME-GAN, we compare with the MBI interpolation \cite{a4}, RBF interpolation \cite{a6}, radioUnet \cite{a8}, auto-encoder (AE) \cite{a9}, deep AE \cite{a9}, traditional Unet, Kriging interpolation \cite{a15}, and traditional cGAN \cite{a30}. The Normalized Mean Square Error (NMSE) and Root Mean Square Error (RMSE) of the reconstructed radio map are present in Table \ref{oa}.

In Setup 1, samples are uniformly distributed. The learning-based approaches have better performance than the model-based and interpolation-based methods since the environment of these regions are more complex and the radio map is sensitive to shadowing effect. Our RME-GAN outperforms all the methods in Setup 1. Especially in comparison with traditional cGAN, our two-phase training process provides more information and display superior performance. We also compare 
different methods over different ratio of observations in Setup 1, shown in Fig. \ref{ratio}. 
Our proposed RME-GAN has the best performance results, as Table \ref{oa} further illustrates.

In Setup 2 and Setup 3, observations are biased among different regions and non-uniformly distributed, respectively. RME-GAN still shows the best results owing to the design of different sampling strategies. The experimental results demonstrate the power of the proposed resampling methods and RME-GAN in radio map reconstruction.

\begin{figure*}[t]
	\centering
	\subfigure[Histogram in a training sample ]{
		\label{hist1}
		\includegraphics[height=4.5cm,width=5.5cm]{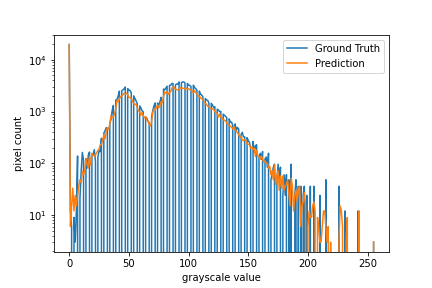}}
	\hspace{4cm}
	\subfigure[Histogram in a test sample]{
		\label{hist2}
		\includegraphics[height=4.5cm,width=5.5cm]{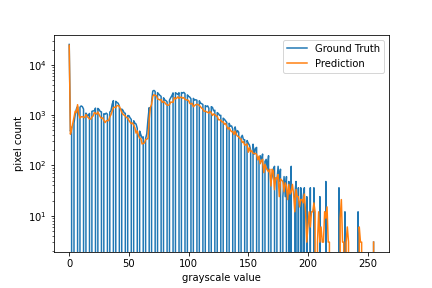}}
	\caption{One Example of Histograms of PSD in Reconstructed Radio Maps}
	\label{ex_hist}
\end{figure*}

Beyond the numerical results obtained, we also present the visualization results of a radio map in Fig. \ref{gt}. The reconstructed results from different methods are shown in Fig. \ref{VISUAL}. Since interpolation-based methods, such as MBI and RBF cannot capture the impact of environment efficiently, it fails to reconstruct the radio map in both three setups. Kriging-based methods provide a blurred construction of radio map as it estimates PSD from neighborhood data. Learning-based methods generally display better reconstruction. Compared to RadioUnet and AE, our RME-GAN constructs more accurate radio maps. Especially in view of Setup 3 where 
the sparse observations are non-uniformly distributed over the region, the proposed RME-GAN outperforms other existing learning methods. 
This happens because our resampling strategy is able to balance the samples over different locations. We also display the histogram of the reconstructed radio map in Setup 1 from RME-GAN as shown in Fig.~\ref{ex_hist}, where RME-GAN is able to capture the distribution of PSD.

\begin{figure}[t]
	\centering
	\includegraphics[height=2.4in, width=2.3in]{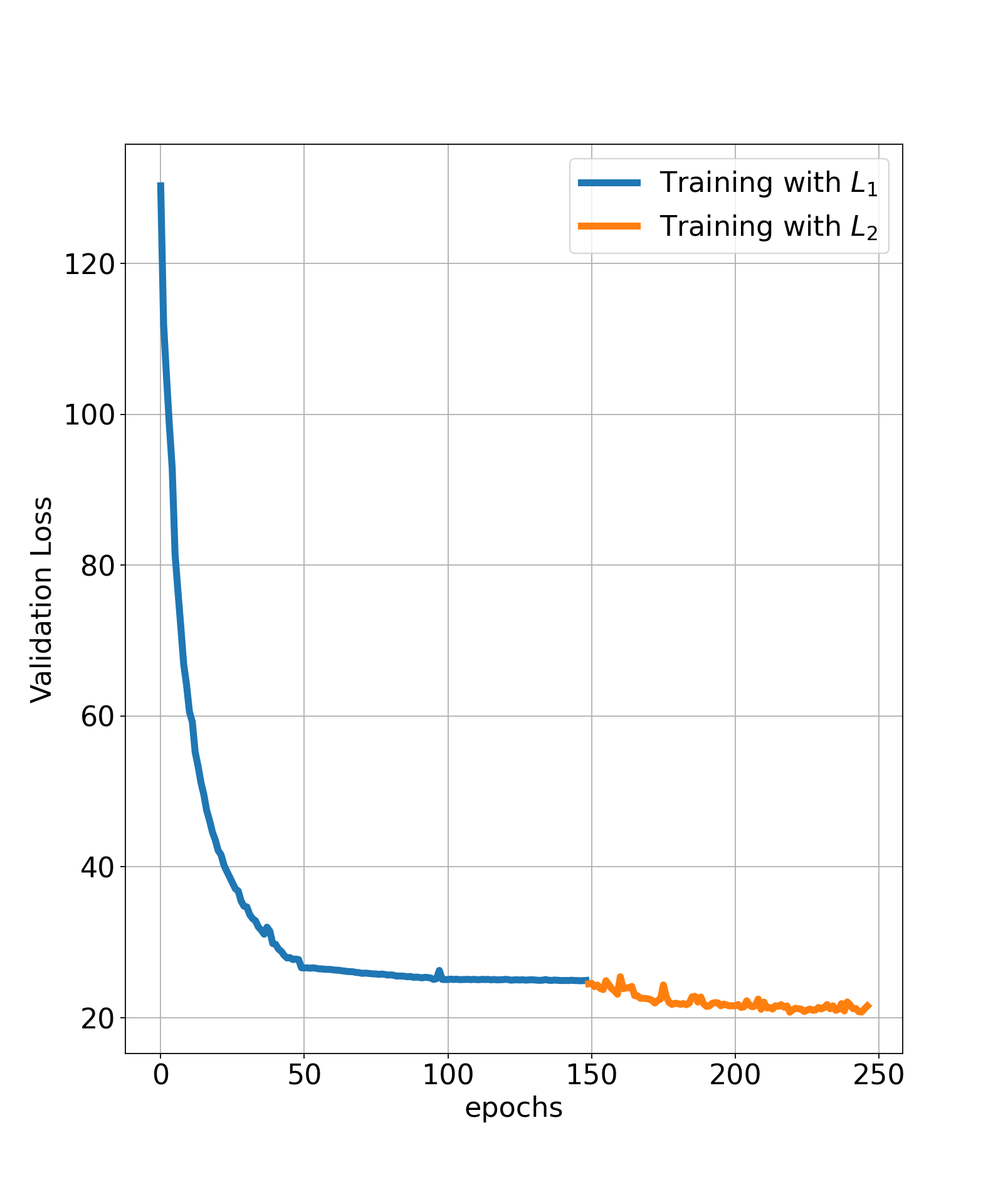}
	\caption{Validation Performance over Different Phase }
	\label{two_phase1}
\end{figure}

\subsection{Robustness}
\subsubsection{Two-Phase Training}
We first confirm the robustness of the two-phase training strategies. 
The validation error over time is shown in Fig. \ref{two_phase1}. From the results, the error drops when the training moves from phase 1 to phase 2, which indicates that the second phase could
better correct detailed errors for the propagation template model from the first phase by leveraging the shadowing information.

\subsubsection{Zero-shot Learning}
Zero-shot learning concerns the recognition of new concepts or new scenarios
when a learning model is trained with generic datasets \cite{a43}. To evaluate the robustness and generalization of the proposed RME-GAN, we train the generator model by using data from Setup 1, but we
test the performance in the data from Setup 2 and 3, respectively. The results are shown in Table \ref{zero-shot}. From the results, we see that our RME-GAN can still obtain a superior performance in the test dataset with different distributions while other learning-based methods all observe a significant performance drop.

\begin{table}[t]
\centering
\caption{Performance of Zero-shot Learning}
\begin{tabular}{|l|ll|ll|}
\hline
\multirow{2}{*}{Model} & \multicolumn{2}{l|}{Setup 2}                           & \multicolumn{2}{l|}{Setup 3}                           \\ \cline{2-5} 
                       & \multicolumn{1}{l|}{NMSE}            & RMSE            & \multicolumn{1}{l|}{NMSE}            & RMSE            \\ \hline
cGAN                   & \multicolumn{1}{l|}{0.1613}          & 0.0918          & \multicolumn{1}{l|}{0.1984}          & 0.1018          \\ \hline
AE                     & \multicolumn{1}{l|}{0.2051}          & 0.1038          & \multicolumn{1}{l|}{0.2259}          & 0.1094          \\ \hline
Deep AE                & \multicolumn{1}{l|}{0.1870}          & 0.0991          & \multicolumn{1}{l|}{0.1981}          & 0.1019          \\ \hline
Unet                   & \multicolumn{1}{l|}{0.0082}          & 0.0208          & \multicolumn{1}{l|}{0.0254}          & 0.0367          \\ \hline
RadioUnet              & \multicolumn{1}{l|}{0.0052}          & 0.0165          & \multicolumn{1}{l|}{0.0148}          & 0.0280          \\ \hline
RME-GAN                 & \multicolumn{1}{l|}{\textbf{0.0042}} & \textbf{0.0150} & \multicolumn{1}{l|}{\textbf{0.0067}} & \textbf{0.0190} \\ \hline
\end{tabular}
\label{zero-shot}
\end{table}

\begin{table}[t]
\centering
\caption{Performance in Selected Areas}
\begin{tabular}{|l|ll|ll|ll|}
\hline
\multirow{2}{*}{Model} & \multicolumn{2}{l|}{Red Area}                          & \multicolumn{2}{l|}{Green Area}                        & \multicolumn{2}{l|}{Blue Area}                         \\ \cline{2-7} 
                       & \multicolumn{1}{l|}{RMSE}            & NMSE            & \multicolumn{1}{l|}{RMSE}            & NMSE            & \multicolumn{1}{l|}{RMSE}            & NMSE            \\ \hline
RBF                    & \multicolumn{1}{l|}{0.0574}          & 0.2794          & \multicolumn{1}{l|}{0.0929}          & 0.3317          & \multicolumn{1}{l|}{0.0868}          & 0.2803          \\ \hline
Kriging                & \multicolumn{1}{l|}{0.0675}          & 0.3867          & \multicolumn{1}{l|}{0.0890}          & 0.3041          & \multicolumn{1}{l|}{0.0819}          & 0.2491          \\ \hline
AE                     & \multicolumn{1}{l|}{0.0545}          & 0.2518          & \multicolumn{1}{l|}{0.0831}          & 0.2651          & \multicolumn{1}{l|}{0.0711}          & 0.1879          \\ \hline
MBI                    & \multicolumn{1}{l|}{0.1087}          & 0.9520          & \multicolumn{1}{l|}{0.1368}          & 0.7183          & \multicolumn{1}{l|}{0.0952}          & 0.3372          \\ \hline
cGAN                   & \multicolumn{1}{l|}{0.0449}                &   0.1714              & \multicolumn{1}{l|}{0.0723}                &     0.2009            & \multicolumn{1}{l|}{0.0466}                &     0.0807            \\ \hline
RadioUnet              & \multicolumn{1}{l|}{0.0164}          & 0.0228          & \multicolumn{1}{l|}{0.0250}          & 0.0240          & \multicolumn{1}{l|}{0.0210}          & 0.0163          \\ \hline
RME-GAN                 & \multicolumn{1}{l|}{\textbf{0.0151}} & \textbf{0.0190} & \multicolumn{1}{l|}{\textbf{0.0221}} & \textbf{0.0189} & \multicolumn{1}{l|}{\textbf{0.0172}} & \textbf{0.0110} \\ \hline
\end{tabular}
\label{3r}
\end{table}

\begin{table}[t]
\centering
\caption{Outage Evaluation (Prediction Error)}
\begin{tabular}{|l|ll|ll|}
\hline
\multirow{2}{*}{Model} & \multicolumn{2}{l|}{Threshod  = 5 }                           & \multicolumn{2}{l|}{Threshod = 25}                           \\ \cline{2-5} 
                       & \multicolumn{1}{l|}{Samples = 1}            & Average & \multicolumn{1}{l|}{Samples = 1}            & Average            \\ \hline
RBF                   & \multicolumn{1}{l|}{0.1165}          & 0.1265         & \multicolumn{1}{l|}{0.2303}          & 0.2349          \\ \hline
MBI                    & \multicolumn{1}{l|}{0.1169}          & 0.1391          & \multicolumn{1}{l|}{0.1599}          & 0.2359         \\ \hline
Kriging                & \multicolumn{1}{l|}{0.1870}          & 0.0991          & \multicolumn{1}{l|}{0.1381}          & 0.1520         \\ \hline
AE                     & \multicolumn{1}{l|}{0.1168}          & 0.1525          & \multicolumn{1}{l|}{0.1498}          & 0.1586          \\ \hline
RadioUnet              & \multicolumn{1}{l|}{0.0085}          & 0.0122          & \multicolumn{1}{l|}{0.0144}          & 0.0158          \\ \hline
RME-GAN                 & \multicolumn{1}{l|}{\textbf{0.0074}} & \textbf{0.0105} & \multicolumn{1}{l|}{\textbf{0.0120}} & \textbf{0.0141} \\ \hline
\end{tabular}
\label{outage-eval}
\end{table}

\subsubsection{Performance in Areas with Lower Energy}
Compared to the areas with larger energy, the areas with lower power spectrum are more sensitive to the environment. To validate the performance of RME-GAN in areas with lower power, we select three areas shown as Fig. \ref{cut} and test different methods in these three regions. The performance is shown in Table \ref{3r}, where RME-GAN outperforms other methods in all areas and display more robustness in the low-power areas.

\subsection{Outage Fault Diagnosis}
We perform outage analysis as a fault diagnosis. 
In some realistic applications, e.g. outage detection, we usually only need a coarse resolution of radio map instead of the fine-resolution one.
To evaluate the power of the proposed RME-GAN in reconstructing radio maps for such applications, we also compare the accuracy of the reconstructed outage maps.
In our definition, 
a location experiences power outage if its PSD falls
below a preset threshold. Then, we can segment the radio map into an outage map with normal and outage areas, i.e., a binary segmentation. We compare the outage maps for all methods against two exemplary thresholds, i.e., 5 and 25. We measure performance by using segmentation errors in Table \ref{outage-eval}. Moreover, we present the results of one randomly selected regions, and the average of 10 randomly selected regions. Our proposed RME-GAN provides a more accurate outage map reconstruction compared to other approaches, which demonstrates the practicability of RME-GAN in network fault detection applications.

\section{Conclusions} \label{Con}
In this work, we propose a two-phase learning frameworks (RME-GAN) for radio map estimation from sparse RF measurements. Our RME-GAN first estimates a template to capture the global radio propagation patterns based model-based interpolation. In the second phase it learns the shadowing effects via geometric and frequency downsampling. The experimental results demonstrate the
performance advantages and the robustness of the proposed RME-GAN, together with the two-phase training strategies in radio map estimation.

The rapid growth and expansion of IoT and 5G systems make efficient estimation and utilization of radio map become increasingly important. Radiomap estimation are important tools for 
resource allocation and network planning.
One promising future direction may be fault diagnosis 
based on the sparse RF observations from deployed sensors or reports from a few users. Another fruitful direction would be the integration of model-based approaches with learning machines. 
Our future works plan  to consider these and other related directions.

\vfill

\end{document}